\newif\ifnotblinded
\notblindedtrue  

\documentclass[a4paper, 11pt]{article}
\usepackage[natbibapa]{apacite}
\usepackage{amsmath}
\usepackage{mathtools}
\usepackage{amssymb}
\usepackage{setspace}
\usepackage{placeins}
\usepackage{minibox}
\usepackage{graphicx}
\usepackage{tikz}
\usepackage{xcolor}
\usepackage{natbib}
\usepackage{caption}
\usepackage{booktabs}
\usepackage{multirow}
\usepackage{authblk}
\usepackage{epstopdf}
\usepackage{algpseudocode}
\usepackage[hmargin = 3cm, vmargin = 2.5cm]{geometry}
\usepackage{lscape}
\usepackage{inputenc}
\usepackage{amsfonts}
\usepackage{latexsym}
\usepackage{float}
\usepackage{siunitx}

\begin{document}
\title{Loss-based Bayesian Sequential Prediction of Value at Risk with a Long-Memory and Non-linear Realized Volatility Model }
\ifnotblinded
\author{Rangika Peiris}
\author{Minh-Ngoc Tran}
\author{Chao Wang\footnote{Corresponding author. Email: chao.wang@sydney.edu.au.}}
\author{Richard Gerlach}

\affil{Discipline of Business Analytics}
\affil{Business School}
\affil{The University of Sydney, Australia}
\fi
\maketitle



\begin{abstract}

A long-memory and non-linear realized volatility model class is proposed for direct Value-at-Risk (VaR) forecasting. This model, referred to as RNN-HAR, extends the heterogeneous autoregressive (HAR) model, a framework known for efficiently capturing long memory in realized measures, by integrating a Recurrent Neural Network (RNN) to handle non-linear dynamics. Loss-based generalized Bayesian inference with Sequential Monte Carlo is employed for model estimation and sequential prediction in RNN-HAR. The empirical analysis is conducted using daily closing prices and realized measures from 2000 to 2022 across 31 market indices. The proposed model's one-step-ahead VaR forecasting performance is compared against a basic HAR model and its extensions. The results demonstrate that the proposed RNN-HAR model consistently outperforms all other models considered in the study.

\vspace{0.5cm}
\end{abstract}


\newpage

\section{Introduction}

Volatility forecasting is a fundamental aspect of financial markets, playing a crucial role for both regulators and market practitioners involved in risk management and asset pricing. Accurate predictions of market volatility are vital for numerous applications, including setting capital reserves, pricing derivatives, and managing investment portfolios. The ability to anticipate market fluctuations can significantly enhance decision-making processes and mitigate financial risks.

Traditionally, parametric models are employed to forecast financial market volatility due to their ease of implementation and interpretability. Among these, the Generalized Autoregressive Conditional Heteroskedasticity (GARCH) model and stochastic volatility models are widely used. The GARCH-type models, first introduced by \cite{bollerslev1986generalized} and further developed by others, provide a robust framework for capturing time-varying volatility by modeling the conditional variance as a function of past variances and returns. These models have proven effective in various financial contexts and remain a staple in the volatility forecasting literature \citep{taylor2008modeling}.

However, traditional models often face challenges when applied to high-frequency intraday data, which is now widely available for many financial assets. Intraday data provides a granular view of market movements, offering richer and more detailed information compared to daily or lower-frequency data \citep{andersen2003modeling}. Despite their popularity, short-memory models like the standard GARCH and stochastic volatility models struggle to accurately capture certain stylized features observed in high-frequency financial data, such as volatility clustering and long-range dependencies \citep{cont2001empirical}. Furthermore, these models typically do not fully leverage the richness of high-frequency data, potentially missing valuable insights.

To address these limitations, researchers have explored long-memory volatility models. Among these, the FIGARCH (Fractionally Integrated GARCH) model introduced by \citet{baillie1996fractionally} has been particularly influential. FIGARCH extends the GARCH framework by incorporating fractional differencing to model long memory in volatility. While effective, these long-memory models often present estimation challenges and lack parsimony, making them less practical for widespread use \citep{tsay2010analysis}. Additionally, despite their advances, they still do not fully utilize high-frequency data.
This gap has led to the development of models that explicitly incorporate high-frequency data to improve volatility forecasting. One such model is the Heterogeneous Autoregressive (HAR) model, introduced by \citet{corsi2009simple}, designed to capture the long memory in realized volatility by leveraging information from high-frequency data.

The HAR model represents a significant advancement in financial econometrics by addressing the limitations of traditional volatility models. It captures long memory in realized volatility measures, making it suitable for analyzing high-frequency financial data. The HAR model operates as an additive cascade model, decomposing volatility into components influenced by different market participants' actions. Although not being formally a long-memory model, the HAR model effectively captures volatility persistence and other stylized facts observed in financial data streams. Its original formulation, using realized variance (RV) and ordinary least squares for estimation, can be extended to address patterns such as non-Gaussianity, spikes/outliers, and conditional heteroskedasticity \citep{clements2021practical}. 
Section \ref{sec:Background models} provides a review of HAR and its extensions.

Despite its strengths, a limitation of the HAR model and its extensions is their reliance solely on realized measures to forecast volatility, ignoring the additional information contained in the return series. This can lead to misleading results, as realized measures can be noisy \citep{hansen2005forecast}. Returns, which encapsulate the cumulative effect of market dynamics, provide valuable insights into investor behavior, market trends, and external factors influencing asset prices. By incorporating returns into volatility modeling, our approach aims to enrich the predictive power of the HAR model. This integration not only reduces the impact of noisy realized measures but also leverages the stability and information inherent in the returns. Ultimately, one of our primary contributions is to extend the HAR model to include returns, thereby enhancing the accuracy and robustness of volatility forecasts, improving risk management practices and contributing to a more comprehensive understanding of financial market dynamics.

While HAR models are effective in predicting realized volatility, applications in financial risk management often necessitate forecasting Value-at-Risk (VaR). VaR is indispensable for assessing and managing financial risk across various contexts, crucial for regulatory compliance and portfolio management strategies, as highlighted in \citet{frey2010quantitative} and \citet{christoffersen2011elements}. Recognizing the practical importance of VaR, our research focuses on directly modeling and forecasting VaR using the HAR framework. This perspective helps  enhance the relevance of HAR models in real-world financial applications. To rigorously evaluate the accuracy of our VaR forecasts, we employ quantile scores, which provide a transparent and robust assessment of predictive performance. 
This methodological choice facilitates a straightforward comparison and validation of our forecasting models, ensuring that our approach meets the stringent demands of risk management practices without making restrictive assumptions about return distributions. Therefore, the second contribution of our research is advancing the utility of HAR models by empowering them to directly forecast VaR and hence avoiding marking assumption on the return distribution, thereby robustifying the volatility modelling and forecasting practice.

Incorporating machine learning (ML) techniques into financial econometrics has become increasingly prominent, offering significant advantages in improving predictive accuracy and modeling intricate relationships within financial data \citep{Kim:2018,nguyen2022long,nguyen2022recurrent}. ML algorithms, such as deep neural networks and random forests, have been successfully deployed in many industrial-level applications. 
Despite the effectiveness of the HAR model and its extensions in capturing volatility dynamics, inherent limitations persist. The HAR model, by design, relies on a linear regression framework, which might restrict its ability to capture complex, non-linear serial dependencies and long-range memory effects inherent in financial time series. Recently, ML methods have gained attention for enhancing HAR model performance.
For example, by comparing the HAR model and Feedforward Neural Network (FNN), \cite{arneric2018neural} develop FNN-HAR models, which are better at capturing the nonlinear behaviour of realized measures and outperform traditional HAR-type models. 
Recognizing these advancements, 
the third contribution of our research makes a significant stride by integrating recurrent neural networks (RNN) into the HAR framework.
More precisely, we derive daily, weekly and monthly effects of realized variances using three RNN structures, thereby capturing the non-linear and long-term effects of these variances on VaR.
We will refer to our approach as RNN-HAR.
It is well-known that RNNs are more efficient than FNNs in terms of capturing serial-dependencies in time series data \citep{lipton2015criticalreview}. 
This integration leverages the ability of RNNs to capture nonlinear dynamics and long-term dependencies in financial data. By embedding RNNs within the HAR model, our methodology aims to improve the accuracy and robustness of VaR predictions. 

Recent advancements in financial econometrics have leveraged sophisticated statistical inference techniques to enhance the accuracy of risk forecasting models. 
Our work utilizes loss-based generalized Bayesian inference, in conjunction with Sequential Monte Carlo (SMC) methods, for model estimation and prediction in RNN-HAR. Loss-based Bayesian inference is invaluable in scenarios where the likelihood function may be challenging to specify or is not readily available; see, e.g., \cite{bissiri2016general} and \cite{knoblauch2019generalized}. 
This approach does not require assumption on the distribution of the returns, avoiding the possible issue of model misspecification.
Given the complexity of our proposed RNN-HAR model structure, using SMC for Bayesian inference and sequential prediction is pivotal, allowing us to handle the inherent challenges of Bayesian computation in sophisticated models such as RNN-HAR.

In summary, the novelty of this research is fourfold.
First, we enrich the HAR framework with the information from the return series. Second, we model and predict VaR directly.
Third, we extend the HAR model by incorporating RNNs.
Fourth, we consider loss-based generalized Bayesian inference with SMC for model estimation and prediction.
Lastly, we evaluate the performance of our proposed model against basic HAR and three other extended HAR models using empirical data spanning nearly two decades (2000–2022) including 31 market indices, demonstrating its superior forecasting capabilities.

This paper is organized as follows. Section 2 reviews the relevant background models. Section 3 proposes the RNN-HAR model. Bayesian inference and prediction using sequential Monte Carlo is presented in section 4. Section 5 presents the empirical results. Section 6 concludes the paper. 

\section{Background models}\label{sec:Background models}
This section presents the HAR model and its extensions. We focus on a selection of widely recognized volatility forecasting models. These models offer diverse methodologies for capturing volatility dynamics and have been extensively studied in the literature for their efficacy in risk management and forecasting applications. We provide a detailed exposition of each model and its respective formulations.

\citet{corsi2009simple} considers the volatility, i.e. the square root of the conditional variance of the return, as generated by several market components in different time horizons. Specifically,  one-day latent partial volatility is denoted as ${\sigma}_{t}^d$, one-week latent partial volatility as ${\sigma}_{t}^w$ and one-month latent partial volatility as ${\sigma}_{t}^m$. Among these, the daily case is the highest frequency volatility component, where ${\sigma}_{t}^d$ is the daily volatility component. The daily return process $y_{t}$ is a function of the daily volatility component. 
\begin{align}
y_{t} &= \sigma_{t}^d\epsilon_{t}, \notag  
\end{align}
where the $\epsilon_{t}$ are i.i.d. with mean 0 and unit variance. Corsi's three-factor stochastic volatility model is based on recursive substitutions of partial volatilities, 
\begin{align}
\sigma _{t+1d}^{d} &= c+ \beta^dRV_{t}^d + \beta^wRV_{t}^w + \beta^mRV_{t}^m + \tilde{\omega}_{t+1d}^d, \label{eq:1}
\end{align}
where $RV_{t}^d$, $RV_{t}^w$ and $RV_{t}^m$ are daily, weekly, and monthly observed realized volatilities respectively. The volatility innovation $\tilde{\omega}_{t+1d}^d$ is contemporaneously and serially independent zero-mean nuisance variate, with an appropriately truncated left tail to guarantee the positive of partial volatilities. From this process for the latent volatility, functional form in terms of realized volatilities can be written as follows, derived for ex-post $\sigma _{t+1d}^{d}$
\begin{align}
\sigma _{t+1d}^{d} &= RV_{t+1d}^d + {\omega}_{t+1d}^d,    \label{eq:2}
\end{align}
where ${\omega}_{t}^d$ subsumes both latent daily volatility measurement and estimation errors. Equation (\ref{eq:2}) links the ex-post volatility estimate $RV_{t+1d}^d$ to the contemporaneous measure of daily latent volatility $\sigma _{t+1d}^{d}$. By substituting Equation (\ref{eq:2}) into (\ref{eq:1}), \citet{corsi2009simple} proposes a time series representation of the cascade model named HAR, where the measurement errors on the dependent variable can be absorbed into the disturbance term of the regression. 
\begin{align}
RV_{t+1d}^{d} = c + \beta^{d}RV_{t}^d + \beta^{w}RV_{t}^w + \beta^{m}RV_{t}^m + {\omega}_{t+1d},   \label{eq:3}
\end{align}
where ${\omega}_{t+1d}  = \tilde{\omega}_{t+1d}^d -{\omega}_{t+1d}^d$.

The HAR model's simplicity allows for various extensions, which can enhance its performance in different ways. One approach is applying transformations to realized volatility, which impacts the model's structure and properties. \cite{corsi2008volatility} describe the square root transformation of the HAR model (SqrtHAR) as follows:
\begin{align}  \label{eq:4}
    \sqrt{RV_{t}} = \alpha_{0} + \alpha_{d}\sqrt{RV_{t-1}} + \alpha_{w}(\sqrt{RV_{t-1}})_{t-5:t-1} + \alpha_{m}(\sqrt{RV_{t-1}})_{t-22:t-1} + u_{t}.
\end{align}
They show that this SqrtHAR model helps stabilize variance and improve the robustness of volatility forecasts compared to the original HAR model \citep{corsi2009simple}. 
Another approach involves modifying the structure of the model to incorporate additional factors. For instance, \cite{corsi2012discrete} introduce the LevHAR model, which integrates past aggregated negative returns into the HAR model to capture the leverage effect. Following \cite{asai2012asymmetry}, we only include the negative part of heterogeneous return since the positive part is usually insignificant. Therefore, we use the following definition of LevHAR model as in \cite{asai2012asymmetry}
\begin{align}  \label{eq:5}
        RV_{t} = \beta_{1}+\beta_{2}RV_{t-1}+ \beta_{3}(RV_{t})_{t-5}+\beta_{4}(RV_{t})_{t-20}+\beta_{5}y_{t-1}I\left [ y_{t-1} < 0\right ]+ ... \\
    \beta_{6}(y_{t})_{t-5}I\left [ (y_{t})_{t-5} < 0\right ]+\beta_{7}(y_{t})_{t-20}I\left [ (y_{t})_{t-20} < 0\right ]+ \textup{error} ,\notag
\end{align}
where $(RV_{t})_{t-h}$ defines as the average of the past $h$ periods' realized variances and $(y_{t})_{t-h}$ is defined in the same manner for return. $I[r<0]$ is the indicator function, which takes 1 if $y_{t}$ is negative and 0 otherwise. 

The Semi-variance HAR (SHAR) model, proposed by \cite{patton2015good}, addresses the asymmetric impact of positive and negative returns on volatility by utilizing daily realized variance split into semi-variances. In addition, the Heterogeneous Autoregressive Quantile (HARQ) model, introduced by \cite{bollerslev2016exploiting}, adjusts coefficients based on measurement errors and necessitates realized quarticity—a metric capturing the fourth moment of realized returns distribution—as a crucial input. Despite their acclaimed capabilities in enhancing volatility estimation and forecasting, our empirical study cannot incorporate these models due to the absence of required input metrics—specifically, daily semi-variance and realized quarticity data—in our dataset. This limitation highlights the challenge of applying these influential models in our empirical analysis.

While the HAR-RV, SqrtHAR, LevHAR, SHAR and HARQ models represent significant advancements in volatility modeling, they share common limitations. These models primarily focus on capturing volatility dynamics through realized volatilities and other factors but do not explicitly incorporate returns into their frameworks. This omission is critical as returns play a fundamental role in determining asset price movements and directly influence risk measures such as VaR. By neglecting to integrate returns and forecast VaR, these models might overlook crucial aspects of financial risk assessment, limiting their applicability in practical risk management contexts. Addressing these gaps is essential for developing more comprehensive and accurate models that better reflect the complexities of financial markets.

Another potential avenue for enhancing modelling accuracy involves integrating the HAR model with the GARCH equation. \cite{huang2016modeling} introduce the Realized-HAR-GARCH model, which expands the volatility dynamic equation by integrating the HAR structure of realized variance into the GARCH equation. In this model, they incorporate multiple lags of a realized measure and employ a measurement equation to make necessary adjustments. They focus on capturing latent volatility associated with inter-day returns rather than the intra-day returns captured by realized measures. The authors argue that their Realized-HAR-GARCH model offers more nuanced dynamics for realized measures than the original HAR model. The Realized-HAR-GARCH model is written as:
\begin{align}  \label{eq:6}
y_{t} &= \mu + \sqrt{h_{t}}z_{t},  \\  \notag
h_{t} &= \omega + \beta h_{t-1} + \gamma_{d} RV^d_{t-1} + \gamma_{w} RV^w_{t-1} + \gamma_{m} RV^m_{t-1}  \\   \notag
RV_{t} &= \xi + \phi h_{t} + \tau_{1}z_{t} + \tau_{2}(z^2_{t}-1)+u_{t},\notag
\end{align}
where $h_{t}$ is the conditional variance, $z_t\sim N(0,1)$, and $u_t\sim N(0, \sigma^2_{u})$ with $z_{t}$ and $u_{t}$ being independent. This model includes multiple lags of $RV^d$. While the Realized-HAR-GARCH model improves the capture of long memory in underlying volatility and offers more accurate multi-period out-of-sample volatility forecasts over various forecast horizons \citep{huang2016modeling}, it does not forecast VaR directly. This limitation highlights a gap that our proposed model aims to fill by incorporating VaR forecasting directly into the framework.

Furthermore, \cite{clements2021practical} explore various aspects to improve the HAR model forecast performance. They consider different estimators, data transformation and combination schemes. There, they consider two weighted least squares schemes and robust regression as estimators, log and square root as transformations, and six combinations of the different estimators and transformations in the empirical study. They conclude that their simple remedies outperform the standard HAR and HARQ forecasts. Further, they suggest estimating model parameters under a loss function coherent with the final application of the forecasts, such as VaR forecasting. This suggestion lays the foundation for our proposed model. 

Despite the advancements made by the HAR model and its aforementioned extensions in capturing volatility dynamics, there remain significant gaps. 
These models do not explicitly incorporate returns into their frameworks and often neglect the direct forecasting of VaR. 

To address these gaps, our research aims to leverage recent advancements by integrating the HAR model with a RNN. This combination seeks to capture future volatility and underlying long-memory behavior more effectively than existing models. By incorporating returns and focusing on VaR forecasting, our proposed RNN-HAR model aims to provide a more comprehensive and accurate approach to financial risk assessment. The following section describes the proposed RNN-HAR model in detail.

\section{The proposed RNN-HAR model}

With advancing technology, RNNs have become increasingly valuable across various industries for their ability to capture complex dependencies and long-term memory within sequential data. This capability has sparked interest among researchers to explore their application in economic modeling \citep{nguyen2022recurrent,bucci2020realized,almosova2023nonlinear,nguyen2022long}.

In this study, we aim to integrate RNNs into the HAR model, leveraging the strengths of both frameworks. The HAR model is well-known for capturing the long memory effects in financial volatility, while RNNs excel in learning intricate patterns and nonlinear dynamics from sequential data. By combining these methodologies, our proposed RNN-HAR model seeks to enhance the accuracy and robustness of volatility forecasting in financial markets.



Utilizing available high-frequency data, we compute realized measures such as Realized Variance (RV) \citep{corsi2009simple}, serving as an estimation of unknown daily volatility, $\sigma _{t}^{2}$. We explore two extensions of Equation (\ref{eq:3}). The first involves employing an RNN to handle each time-horizon RV individually. The second extension entails direct estimation of VaR, defined as the $\alpha$-level quantile of the return distribution, for some $\alpha\in(0,1)$. For parametric models such as GARCH, estimating VaR requires an assumption on the distribution of the returns. It is more statistically robust if one can avoid making such an assumption. Our approach models VaR directly and uses the quantile score as the loss function for model estimation, thus bypassing the need for a distributional assumption of the returns.

Let $y_{t}$ denote the return. \cite{fissler2016higher} demonstrate that the forecast $\textup{VaR}_t^{\alpha}$ minimizes the expectation of the quantile score (QS), defined as
\begin{align}
\textup{QS}_{t} &= \left ( y_{t}-\textup{VaR}_{t}^{\alpha}\right )\left ( \alpha-I(y_{t}<\textup{VaR}_{t}^{\alpha})\right ) \label{eq:7}
\end{align}
where the expected quantile score is minimized at the true VaR, making QS the objective function we optimize.

In Equation (\ref{eq:3}), the HAR model relies only on realized measures and ignores the return series. This might be problematic because realized measures can be noisy, excluding overnight variation when markets are closed \citep{hansen2005forecast}. Relying solely on realized measures for risk estimation can thus lead to misleading results. By using the quantile score as the loss function, we incorporate returns directly, aligning with the ultimate goal of volatility modeling, namely computing VaR.

In our proposed model, we express $\textup{VaR}_{t+1}$ as a function comprising an intercept and three distinct RNN structures: one for daily data, another for weekly data, and a third for monthly data. This approach allows us to account for the non-linear impacts that daily, weekly, and monthly realized measures exert on VaR and incorporates long-term memory effects. Specifically, the RNN-HAR model integrates these three different RNN structures tailored to handle daily, weekly, and monthly realized measures as follows:
\begin{align}  
\textup{VaR}_{t+1}^{\alpha}&=\beta_{0}+ \beta_{1}\textup{h}_{t+1}^d+\beta_{2}\textup{h}_{t
+1}^w+\beta_{3}\textup{h}_{t+1}^m,  \label{eq:8} \\  
\textup{h}_{t+1}^d &= \textup{RNN}(RV_{t}^d,\textup{h}_{t}^d)=\phi(\alpha_{0}^d+\alpha_{1}^dRV_{t}^d+\alpha_{2}^d\textup{h}_{t}^d), \label{eq:9}  \\  
\textup{h}_{t+1}^w &= \textup{RNN}(RV_{t}^w,\textup{h}_{t}^w)=\phi(\alpha_{0}^w+\alpha_{1}^wRV_{t}^w+\alpha_{2}^w\textup{h}_{t}^w),  \label{eq:10} \\
\textup{h}_{t+1}^m &= \textup{RNN}(RV_{t}^m,\textup{h}_{t}^m)=\phi(\alpha_{0}^m+\alpha_{1}^mRV_{t}^m+\alpha_{2}^m\textup{h}_{t}^m).  \label{eq:11}    
\end{align} 
Here, the observed realized measures are denoted as ${RV_{1},...RV_{T}}$, with $T$ the total number of observations in the training sample. These data are used to compute daily, weekly, and monthly realized measures as inputs for our analysis. The RNN structure employed in this paper is the simple RNN, with $\phi$ the tanh activation function. 
It is possible to consider more sophisticated RNN structure such as the Long short-term memory model, but we do not consider it here.
The loss function utilized here aggregates the quantile scores across the training period, enabling the computation of $\textup{VaR}_{t+1}$ as a function of daily, weekly, and monthly realized measures using Equation (\ref{eq:8}). This approach ensures comprehensive consideration of the non-linear effects exerted by these realized measures on VaR, thereby enhancing the model's accuracy and reliability. 

Our proposed RNN-HAR model addresses several significant gaps in existing volatility models. Firstly, it integrates RNNs into the HAR framework, allowing for the capture of complex, non-linear dependencies and long-range memory in volatility forecasting. Secondly, by employing the quantile loss function, the model avoids the need to assume specific distributions for returns, which is a common limitation in parametric models like GARCH. Thirdly, we directly forecast VaR rather than RV as in the HAR model, providing a more holistic framework for risk assessment. These contributions underscore the model's potential to significantly enhance volatility forecasting accuracy and risk management strategies in financial markets. The next section will delve into the specifics of model estimation and VaR senquential forecast using the loss-based Bayesian SMC method.

\section{Loss-based Bayesian inference and prediction} 

Loss-based generalized Bayesian inference represents a paradigm shift in statistical modeling, diverging from traditional methods that rely on likelihood functions and strict distributional assumptions. 
See, e.g., \cite{bissiri2016general,knoblauch2019generalized,matsubara2021robust} and \cite{frazier2024loss}.
Unlike classical frameworks where likelihood functions necessitate specific probabilistic assumptions about the data distribution, loss-based Bayesian inference emphasizes the use of loss functions to guide Bayesian inference and prediction. This approach is particularly advantageous in scenarios where underlying data distributions are complex or unknown, mitigating potential model misspecification by offering flexibility and robustness. By focusing on a loss function rather than a likelihood, researchers can tailor models to better reflect real-world uncertainties and variations by updating beliefs about model parameters in a robust manner, particularly in situations where justifying strict assumptions about data distributions is challenging.
Recent developments, as discussed in \cite{bissiri2016general} and \cite{knoblauch2019generalized}, highlight the use of loss functions in updating beliefs and parameter estimation without restrictive distributional assumptions, thereby enhancing the applicability and reliability of Bayesian models in complex data environments.

Building on these principles, we adopt a loss-based Bayesian approach to address the challenge of modeling financial volatility and forecasting VaR without presuming a specific return distribution. This method effectively integrates prior knowledge with observed data, facilitating robust parameter estimation without imposing strong distributional assumptions.

In the volatility modeling literature, \cite{taylor2019forecasting}  advocates for methods that estimate parameters using the quantile score loss function. 
Using this quantile loss function leads to the following 
asymmetric Laplace (AL) distribution on the return $y_t$,
\begin{equation}\label{eq:AL dist}
f(y_{t}|Q_t,\sigma)=\frac{\alpha(1-\alpha)}{\sigma}\textup{exp}\big(-(y_{t}-Q_{t})(\alpha-I(y_{t}\leqslant Q_{t}))/\sigma\big)
\end{equation}
where $Q_{t}$ denotes the time-varying location parameter representing the quantile corresponding to the chosen probability level $\alpha$, and $\sigma$ is the scale parameter.
Our model utilizes the AL distribution \eqref{eq:AL dist} for generalized Bayesian inference and prediction.   
We adopt an inverse Gamma prior on $\sigma$, leading to its full conditional distribution being inverse Gamma, thus allowing straightforward integration and simplifying the likelihood function in subsequent Bayesian analysis steps. As a result, the posterior distribution of the parameter of interest {\( \theta \) is obtained without explicit consideration of 
$\sigma$ \citep{gerlach2011bayesian}.

In Bayesian inference, priors serve as our initial assumptions regarding model parameters before observing any data. 
Following an exploration of different prior combinations, we opted for a normal prior with a mean of zero and a variance of 0.01 for the recurrent parameters in our RNN-HAR model. This choice reflects our initial expectation that these parameters are centered around zero with small variability. For the model parameters $\beta_i, i=1,...,4$, we assumed a normal distribution with a mean of 0 and a standard deviation of 1, indicating our initial uncertainty and allowing for exploration across a spectrum of parameter values.

Integrating loss-based generalized Bayesian inference into the RNN-HAR framework yields several advantages: providing a coherent methodology for updating beliefs based on observed data, 
robustifing inference for the model parameters without relying on data assumptions of the returns, allowing convenient and efficient sequence prediction based on Sequence Monte Carlo.



\subsection{Sequential Monte Carlo (SMC)} 

The SMC method is an attractive approach for Bayesian inference and sequential prediction; see, e.g., \cite{del2006sequential} and \cite{gunawan2022flexible}.
SMC uses a set of samples, often called particles, to approximate a sequence of probability distributions.
This sequential updating enables efficient estimation of posterior distributions in complex and non-linear models where conventional Monte Carlo methods are computationally demanding or impractical. 
SMC allows straightforward expanding-window one-step-ahead forecast calculations, which makes it particularly useful for volatility forecasting \citep{nguyen2022recurrent}.
The SMC method also provides an accurate estimate of the marginal likelihood, which is an important quality often used for model selection.
These attributes make SMC an attractive approach for Bayesian inference and sequential forecasting in our RNN-HAR model. 

There are two common SMC approaches in the literature: likelihood annealing and data annealing \citep{nguyen2022recurrent}. 
The first approach is designed for sampling from the posterior while the second is for sequential prediction. 
We present these two approaches in the next sections.

\subsubsection{Likelihood annealing}
The loss-based generalized posterior distribution in our RNN-HAR model is 
\begin{equation}
 \pi (\theta)=p(\theta|y_{1:T})\propto p(y_{1:T}|\theta)p(\theta),  \label{eq:13}
\end{equation}
 where $p(\theta)$ is the prior and $p(y_{1:T}|\theta)$ is the loss-based likelihood-alike function derived from \eqref{eq:AL dist}.
For sampling from the generalized posterior $\pi(\theta)$, SMC first samples a set of $M$
weighted particles $\left \{ {W_{0}^{j},\theta_{0}^{j}} \right \}_{j=1}^{M}$ from an easy-to-sample distribution $\pi_{0}(\theta)$, such as the prior $p(\theta)$, and then traverses these particles through intermediate distributions $\pi_{t}(\theta), t=1,...,K$ with $\pi_{K}(\theta)=\pi(\theta)$.
In this paper, we set $\pi_{0}(\theta) = p(\theta)$ as it is possible to sample from the prior $p(\theta)$. 
The likelihood annealing SMC sampler uses the following intermediate distributions
\begin{equation}  
    \pi_{t}(\theta):= \pi_{t}(\theta|y_{1:T}) \propto p(y_{1:T}|\theta)^{\gamma_{t}}p(\theta),   \label{eq:14}
\end{equation}
where the $\gamma_{t}$ are referred to as the temperature levels satisfying $0 = \gamma_{0}<\gamma_{1}<\gamma_{2}<...<\gamma_{k}=1$.  
Note that the sequence of distributions $\pi_t$ requires the full training data $y_{1:T}$ to be available. 
SMC with a likelihood annealing sampler is suitable for in-sample analysis.

Several methods exist to implement SMC in practice; here, we consider the method used by \citet{nguyen2022recurrent}, which uses three main steps: reweighting, resampling, and a Markov move. \newline \newline
\textbf{Reweighting}: At the beginning of iteration $t$, the set of weighed particles $\left \{ {W_{t-1}^{j},\theta_{t-1}^{j}} \right \}_{j=1}^{M}$ that approximate the intermediate distribtuion $\pi_{t-1}(\theta)$ is reweighted to approximate the target $\pi_{t}(\theta)$. The efficiency of these weighted particles as a representation of $\pi_{t}(\theta)$ is often measured by the effective sample size (ESS) defined in \eqref{eq:17}.  \newline \newline
\textbf{Resampling}: The particles are resampled if the ESS is below a prespecified threshold. \newline \newline
\textbf{Markov Move}: The resulting equally weighted samples are then refreshed by a Markov kernel whose invariant distribution is $\pi_{t}(\theta)$. \\

Following \citet{nguyen2022recurrent}, we choose the tempering sequence $\gamma_t$ adaptively to ensure a sufficient level of particle efficiency by selecting the next value of $\gamma_t$ such that ESS stays above a threshold. We now present the likelihood annealing SMC sampler, which is adapted from \citet{nguyen2022recurrent}.
Note that we do not compute the marginal likelihood estimate as its meaning is not well justified in the generalized Bayesian inference setting.
\newline 
\begin{enumerate}
        \item Sample $\theta_{0}^j\sim p(\theta) $ and set $W_{0}^j= 1/M$ for $j=1...M$
        \item For $t=1,...,K$
        \begin{enumerate}
            \item Resampling: Compute the unnormalized weights
            \begin{equation}
                w_{t}^{j} = W_{t-1}^{j}\frac{p(y_{1:T}|\theta^j_{t-1})^{\gamma_t}p(\theta^{j}_{t-1})}{p(y_{1:T}|\theta^j_{t-1})^{\gamma_{t-1}}p(\theta^{j}_{t-1})}= W_{t-1}^{j}p(y_{1:T}|\theta_{t-1}^j)^{\gamma_t-\gamma_{t-1}}, j=1,...,M    \label{eq:15}
            \end{equation}
                and set the new normalized weights
                \begin{equation}
                w_{t}^{j} = \frac{w_{t}^{j}}{\sum_{s=1}^{M}w_{t}^{s}}, j = 1,...,M  \label{eq:16}
               \end{equation}

            \item Compute the effective sample size (ESS)
            \begin{equation}
            ESS = \frac{1}{\sum_{j=1}^{M}\left ( w_{t}^{j}\right )^2}, j = 1,...,M  \label{eq:17}
            \end{equation}
        \textbf{if} $\textup{ESS} < cM$ for some $0 < c < 1$, \textbf{then}
        \begin{itemize}
            \item Resampling: Resample from $\left \{ \theta^j_{t-1}\right \}^M_{j=1}$ using the weights  $\left \{W^j_{t}\right \}^M_{j=1}$, and then set ${W^j_{t}}=1/M$ for $j = 1,...,M $, to obtain the new equally-weighted particles $\left \{ \theta^j_{t},W^j_{t}\right \}^M_{j=1}$.
            \item Markov move: For each $j=1,...,M$ move the sample $\theta_{t}^j$ according to $N_\text{lik}$ random walk Metropolis-Hasting steps:
            \begin{itemize}
                \item Generate a proposal $\theta_{t}^{{j}'}$ from a from a multivariate normal distribution $\textit{N}(\theta^j_{t},\Sigma_{t})$ with $\Sigma_{t}$ the covariance matrix.
                \item Set $\theta^j_{t} = \theta_{t}^{{j}'}$ with the probability
                \begin{equation}
                    \textup{min}\left ( 1,\frac{p(y_{1:T}|\theta_{t}^{{j}'})^{\gamma_t}p(\theta_{t}^{{j}'})}{p(y_{1:T}|\theta^j_{t})^{\gamma_{t}}p(\theta^{j}_{t})}\right );  \label{eq:18}              
                \end{equation}
                otherwise keep $\theta_{t}^j$ unchanged.
            \end{itemize}
        \end{itemize}
        \textbf{end}
   
        \end{enumerate}
\end{enumerate}
\subsubsection{Data annealing}
For out-of-sample expanding-window forecasts where the posterior of the model parameters $\theta$ is updated once new data arrive, it is necessary to use SMC with the data annealing \citep{nguyen2022recurrent}. The following sequence of distributions is used to generate weighted particles in this SMC sampler. 
\begin{equation}
\pi_{t}(\theta):= \pi_{t}(\theta|y_{1:t}) \propto p(y_{1:t}|\theta)p(\theta)\propto \pi_{t-1}(\theta)p(y_{t}|\theta,y_{1:t-1}),\;\;t=T+1,...  \label{eq:20}
\end{equation}
with $y_{1:t}$ the data available up to time $t$, and $y_{1:T}$ the in-sample data. 
The SMC procedure for sampling from the sequence $\pi_{t}(\theta)$ in \eqref{eq:20} is the same as before, except that the 
unnormalized weights become
\begin{equation}
    w_{t}^{j} = W_{t-1}^{j}\frac{p(y_{1:t}|\theta^j_{t-1})p(\theta^{j}_{t-1})}{p(y_{1:t-1}|\theta^j_{t-1})p(\theta^{j}_{t-1})}= W_{t-1}^{j}p(y_{t}|y_{1:t-1,}\theta^j_{t-1}), j=1,...,M    \label{eq:21}
\end{equation}
In line with \citet{nguyen2022recurrent}, we employ SMC with likelihood annealing for in-sample Bayesian inference and SMC with data annealing for generating one-step-ahead forecasts. The specific implementation settings for the SMC samplers are outlined below.

\begin{table}[H]
\caption{SMC settings}  \label{table:1}
\begin{tabular}{|l|l|l|}
\hline
Variable   & Description                                                 & Value  \\ \hline
$K$          & Number of annealing levels                                  & 10,000 \\ \hline
$M$          & Number of particles                                         & 2,000 \\ \hline
$c$          & Constant of the ESS threshold                               & 0.8    \\ \hline
$N_\text{lik}$  & Number of Markov moves in the SMC with likelihood annealing & 10     \\ \hline
$N_\text{data}$ & Number of Markov moves in the SMC with data annealing       & 20     \\ \hline
\end{tabular}
\end{table}

\section{Data and Empirical study} 
The daily closing prices and realized measure data utilized in this study were sourced from the Oxford-man Institute’s realized library \citep{heber2009omi}, covering the period from 2000 to 2022. Daily return values were computed based on the daily price data. Our analysis encompasses significant events such as the global financial crisis and the COVID-19 pandemic. Due to varying non-trading days across different markets throughout the period under study, sample sizes and forecasting periods vary across each series.
To ensure consistency, we standardized the length of all return series to the last 3000 observations, except for the BVLG time series, which has only 2398 values. These series were then divided into an in-sample period comprising the first 2000 observations and an out-of-sample period comprising the last 1000 observations. Table \ref{table:2} presents details regarding each market considered. Figure \ref{Figure 1} displays the time series plot of the absolute value of daily return and RV of SPX as an example.

\begin{figure}[H]
\centering
\includegraphics[width=0.9\textwidth]{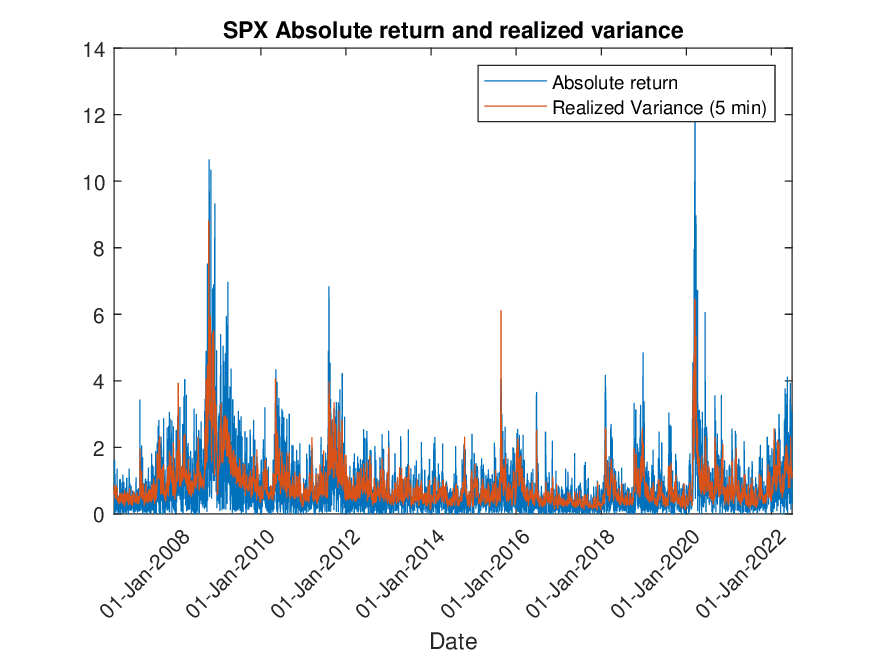}
\caption{SPX absolute return series and realized variance}
\label{Figure 1}
\end{figure}

Table \ref{table:2}  provides descriptive statistics for each market in this study. Among these markets, the emerging market BVSP exhibits the highest standard deviation, indicating greater variability than the more established markets. On the other hand, IXIC stands out as having the highest mean return.
\begin{table}[H]
\caption{Summary statistics of the return series.}\label{table:2}
\centering{}%
\begin{tabular}{lccccccccc}
\hline 
Market & N & Mean & Std & Skewness & Kurtosis & Min & Max\tabularnewline
\hline 
AEX & 3000 & 0.0234 & 1.1183 & -0.5723 & 9.7439 & -10.6384 & 8.6462\tabularnewline
AORD & 3000 & 0.0005 & 0.9052 & -0.8246 & 8.8584 & -7.0728 & 4.4291\tabularnewline
BFX & 3000 & 0.0103 & 1.1379 & -0.9672 & 15.2145 & -14.2235 & 6.8949\tabularnewline
BSESN & 3000 & -0.0081 & 1.1011 & -0.9482 & 16.5074 & -13.8222 & 8.2043\tabularnewline
BVLG & 2398 & 0.0000 & 1.1353 & -0.8747 & 10.1403 & -10.9619 & 6.5051\tabularnewline
BVSP & 3000 & -0.0197 & 1.5750 & -0.7933 & 14.5011 & -16.0260 & 12.9906\tabularnewline
DJI & 3000 & 0.0209 & 1.0794 & -0.9663 & 24.5143 & -13.8247 & 10.7360\tabularnewline
FCHI & 3000 & 0.0156 & 1.2663 & -0.5953 & 9.9544 & -11.9977 & 7.7889\tabularnewline
FTMIB & 3000 & 0.0011 & 1.5565 & -1.0556 & 14.1095 & -18.5434 & 8.5472\tabularnewline
FTSE & 3000 & 0.0089 & 1.0202 & -0.5866 & 10.4931 & -10.1382 & 7.7806\tabularnewline
GDAXI & 3000 & 0.0136 & 1.2752 & -0.4607 & 9.4086 & -11.8749 & 9.7516\tabularnewline
GSPTSE & 3000 & -0.0004 & 0.9093 & -1.8510 & 33.5737 & -13.1944 & 9.1019\tabularnewline
HSI & 3000 & -0.0038 & 1.2255 & -0.1899 & 6.0910 & -5.9839 & 8.7032\tabularnewline
IBEX & 3000 & -0.0026 & 1.3829 & -0.7636 & 11.5616 & -12.7119 & 8.1168\tabularnewline
IXIC & 3000 & 0.0358 & 1.2680 & -0.7839 & 11.9963 & -13.1586 & 8.9088\tabularnewline
KS11 & 3000 & -0.0036 & 1.0185 & -0.5060 & 10.6059 & -10.1935 & 7.0957\tabularnewline
KSE & 3000 & -0.0148 & 1.0454 & -0.5878 & 6.8834 & -7.3188 & 4.6228\tabularnewline
MXX & 3000 & -0.0203 & 0.9770 & -0.5030 & 7.5941 & -6.9709 & 5.0541\tabularnewline
N255 & 3000 & 0.0266 & 1.3326 & -0.4221 & 8.0216 & -11.1593 & 7.7255\tabularnewline
NSEI & 3000 & -0.0014 & 1.1074 & -0.9426 & 15.5131 & -13.6741 & 8.0057\tabularnewline
OMXC20 & 3000 & 0.0109 & 1.1455 & -0.3443 & 5.7130 & -7.8569 & 5.1068\tabularnewline
OMXHPI & 3000 & 0.0102 & 1.1816 & -0.6401 & 8.8448 & -10.7945 & 6.1853\tabularnewline
OMXSPI & 3000 & 0.0056 & 1.1507 & -0.7847 & 10.3512 & -11.8285 & 6.9907\tabularnewline
OSEAX & 3000 & -0.0011 & 1.1294 & -0.6811 & 8.9272 & -9.8730 & 5.8014\tabularnewline
RUT & 3000 & 0.0120 & 1.4329 & -0.8890 & 13.8693 & -15.2513 & 8.8834\tabularnewline
SMSI & 3000 & -0.0041 & 1.3624 & -0.7651 & 12.1109 & -14.0552 & 8.1712\tabularnewline
SPX & 3000 & 0.0250 & 1.1006 & -0.8630 & 18.0908 & -12.6874 & 8.9440\tabularnewline
SSEC & 3000 & -0.0113 & 1.3264 & -0.9804 & 9.5522 & -8.8919 & 5.6243\tabularnewline
SSMI & 3000 & 0.0117 & 0.9885 & -0.8894 & 12.5799 & -10.1410 & 6.7734\tabularnewline
STI & 3000 & 0.0223 & 0.9610 & -1.5480 & 26.6530 & -14.7222 & 5.9946\tabularnewline
STOXX50E & 3000 & 0.0135 & 1.2777 & -0.5419 & 9.6881 & -11.9999 & 8.6605 \tabularnewline
\hline 
\end{tabular}
\end{table}

\subsection{One step forecasting}
We implement the following procedure for one-step forecasting. 
At the time $t$ within the test data ($t \geqslant T+1$),
let $\{\theta^{(i)}\}_{i=1}^M$ be the particles approximating the posterior distribution $\pi_t(\theta)=p(\theta|y_{1:t})$.
For each particle $\theta^{(i)}$, we compute $Q_{t+1}^{(i)}$ according to Eq. \eqref{eq:8}, which represents an estimate of $\text{VaR}_{t+1}^\alpha$.  
The resulting $M$ values $\{Q_{t+1}^{(i)}\}_{i=1}^M$ represent the posterior predictive distribution of $\text{VaR}_{t+1}^\alpha$, given the information up to time $t$.
The arithmetic mean, $\widehat{Q}_{t+1}$, of these realizations serves as the point forecast for $\text{VaR}_{t+1}^\alpha$.
The predictive quantile score, evaluated on the test data, is computed as
\begin{align}  \label{eq: 26}
    QS = \frac{1}{T_{test}}\sum\big(y_{t+1}-\widehat{Q}_{t+1})(\alpha-I(y_{t+1}\leqslant \widehat{Q}_{t+1} )\big).
\end{align} 
We perform an empirical analysis to compare the proposed RNN-HAR model with the conventional HAR model and its extensions. We consider the extensions that can be executed using the available dataset for this comparative analysis, including sqrtHAR as detailed in Equation (\ref{eq:4}), LevHAR as presented in Equation (\ref{eq:5}) and RHARGARCH as described in Equation (\ref{eq:6}).
It is worth noting that except for our proposed RNN-HAR model that directly outputs VaR forecasts, all the other models in our analysis forecast the realized measure.
Except for the RHARGARCH model where the standard equation to calculate VaR can be used, 
we follow \cite{clements2021practical} and calculate VaR from RV as follows
\begin{align}
    \text{VaR}_{t}^{\alpha} = \mu_{t}+\Phi ^{-1}(\alpha)\sqrt{F_{t}},  \label{eq:27}
\end{align}
where $\mu_{t}$ is the conditional mean of the return and $\Phi ^{-1}$ is the inverse of the standard normal cdf and $F_{t}$ denotes a forecast of $RV_{t}$. 

\subsection{Evaluating VaR performance}
To assess the performance of VaR forecasts, we consider several criteria as discussed below. 
Following the guidelines outlined in the Basel III Capital Accord \citet{de2019minimum}, our analysis focuses on daily one-step-ahead VaR forecasts at the quantile level of $\alpha = 2.5\%$. Additionally, we also explore additional quantile levels of $\alpha = 1\%$ and $\alpha = 5\%$ for empirical analysis purposes. 
We employ four predictive measures to assess the performance of VaR forecasting. The primary performance metric is the quantile score defined in Equation (\ref{eq: 26}). A model with the lowest quantile loss is considered preferable. Table \ref{table:3} displays the quantile score values across 31 markets for the three different $\alpha$ values investigated in this study.

The results presented in Table \ref{table:3} indicate the superior performance of the RNN-HAR model. Specifically, for $\alpha = 1\%$, more than 77\% of the markets favour the proposed model, while for $\alpha = 2.5\%$, this figure is 71\%, and for $\alpha = 5\%$, it exceeds 51\%. These findings underscore the effectiveness of the proposed model across varying quantile levels.

\setlength{\tabcolsep}{2pt}
\begin{landscape}
\mbox{}\vspace*{0.3cm} 
\begin{table}[H]
\mbox{}\hspace*{0.25cm} 
\caption{Quantile score values} 
\centering{}{\tiny{}} \label{table:3}     
\scriptsize\begin{tabular}{|l|ccccc|ccccc|ccccc|}  
\multicolumn{1}{l}{} & \multicolumn{5}{c}{$\alpha = 1\%$}   & \multicolumn{5}{c}{$\alpha = 2.5\%$} & \multicolumn{5}{c}{$\alpha = 5\%$}                                                        \\   \hline

Market  & RNN-HAR         & HAR    & LevHAR          & SqrtHAR         & RHARGARCH & RNN-HAR         & HAR             & LevHAR          & SqrtHAR         & RHARGARCH & RNN-HAR         & HAR             & LevHAR          & SqrtHAR         & RHARGARCH       \\    \hline
AEX                  & \textbf{0.0397} & 0.0439 & 0.0504          & 0.0431          & 0.0523    & \textbf{0.0840} & 0.0882          & 0.0887          & 0.0867          & 0.0947    & \textbf{0.1408} & 0.1448          & 0.1473          & 0.1431          & 0.1499          \\
AORD                 & \textbf{0.0380} & 0.0426 & 0.0516          & 0.0420          & 0.0519    & \textbf{0.0756} & 0.0790          & 0.0792          & 0.0782          & 0.0868    & 0.1254          & 0.1243          & 0.1336          & \textbf{0.1232} & 0.1314          \\
BFX                  & \textbf{0.0452} & 0.0482 & 0.0511          & 0.0480          & 0.0551    & \textbf{0.0873} & 0.0925          & 0.0924          & 0.0913          & 0.0982    & \textbf{0.1468} & 0.1503          & 0.1504          & 0.1486          & 0.1554          \\
BSESN                & \textbf{0.0460} & 0.0505 & 0.0711          & 0.0492          & 0.0484    & \textbf{0.0871} & 0.0900          & 0.0903          & 0.0880          & 0.0880    & 0.1424          & 0.1415          & 0.1563          & \textbf{0.1391} & 0.1397          \\
BVLG                 & \textbf{0.0404} & 0.0422 & 0.0436          & 0.0663          & 0.0432    & \textbf{0.0826} & 0.1079          & 0.0831          & 0.1067          & 0.0839    & 0.1377          & 0.1386          & \textbf{0.1375} & 0.1600          & 0.1395          \\
BVSP                 & 0.0682          & 0.0644 & 0.064           & \textbf{0.0631} & 0.0634    & 0.1169          & 0.1167          & \textbf{0.1165} & 0.1175          & 0.1172    & 0.1924          & \textbf{0.1876} & 0.1901          & 0.1892          & 0.1893          \\
DJI                  & \textbf{0.0436} & 0.0508 & 0.0565          & 0.0490          & 0.0515    & \textbf{0.0834} & 0.0912          & 0.0919          & 0.0887          & 0.0920    & \textbf{0.1389} & 0.1461          & 0.1504          & 0.1420          & 0.1446          \\
FCHI                 & \textbf{0.0477} & 0.0552 & 0.0578          & 0.0545          & 0.0574    & \textbf{0.0953} & 0.1013          & 0.1014          & 0.0996          & 0.1029    & \textbf{0.1544} & 0.1610          & 0.1616          & 0.1587          & 0.1617          \\
FTMIB                & \textbf{0.0643} & 0.0667 & 0.0665          & 0.0664          & 0.0678    & 0.1145          & 0.1145          & 0.1144          & \textbf{0.1142} & 0.1158    & \textbf{0.1746} & 0.1757          & 0.1753          & 0.1753          & 0.177           \\
FTSE                 & \textbf{0.0440} & 0.0501 & 0.0517          & 0.0486          & 0.0593    & \textbf{0.0861} & 0.0900          & 0.0912          & 0.0886          & 0.0987    & \textbf{0.1383} & 0.1396          & 0.1407          & 0.1384          & 0.1471          \\
GDAXI                & \textbf{0.0528} & 0.0649 & 0.0641          & 0.0634          & 0.0600    & \textbf{0.0990} & 0.1101          & 0.1101          & 0.1082          & 0.1051    & \textbf{0.1601} & 0.1679          & 0.1668          & 0.1660          & 0.165           \\
GSPTSE               & \textbf{0.0426} & 0.0507 & 0.0569          & 0.0483          & 0.0563    & \textbf{0.0774} & 0.0803          & 0.08            & 0.0789          & 0.0872    & \textbf{0.1134} & 0.1195          & 0.127           & 0.1188          & 0.1266          \\
HIS                  & \textbf{0.0426} & 0.0471 & 0.0427          & 0.0462          & 0.0481    & \textbf{0.0919} & 0.0934          & 0.0937          & 0.0926          & 0.0948    & \textbf{0.1550} & 0.1563          & 0.1552          & 0.1556          & 0.1568          \\
IBEX                 & 0.0517          & 0.0506 & \textbf{0.0499} & 0.0500          & 0.0569    & 0.0961          & \textbf{0.0938} & 0.0941          & 0.0939          & 0.1015    & 0.1509          & 0.1512          & 0.1509          & \textbf{0.1506} & 0.1564          \\
IXIC                 & \textbf{0.0464} & 0.0484 & 0.0504          & 0.0472          & 0.0484    & 0.0993          & 0.0996          & 0.0996          & \textbf{0.0985} & 0.0993    & 0.1675          & 0.1695          & 0.1721          & \textbf{0.1669} & 0.1684          \\
KS11                 & 0.0367          & 0.0398 & \textbf{0.036}  & 0.0382          & 0.0393    & \textbf{0.0741} & 0.0780          & 0.0779          & 0.0758          & 0.0777    & 0.1274          & 0.1292          & \textbf{0.1251} & 0.1266          & 0.1283          \\
KSE                  & \textbf{0.0399} & 0.0489 & 0.0464          & 0.0488          & 0.0460    & \textbf{0.0836} & 0.0901          & 0.0901          & 0.0897          & 0.0872    & 0.1422          & 0.1445          & 0.1422          & 0.1438          & \textbf{0.1419} \\
MXX                  & 0.0427          & 0.0438 & \textbf{0.0392} & 0.0423          & 0.0465    & 0.0809          & 0.0806          & 0.0803          & \textbf{0.0795} & 0.0827    & 0.1324          & 0.1297          & \textbf{0.1276} & 0.1292          & 0.1307          \\
N225                 & \textbf{0.0391} & 0.0412 & 0.0439          & 0.0405          & 0.0446    & \textbf{0.0810} & 0.0836          & 0.0838          & 0.0827          & 0.0877    & \textbf{0.1398} & 0.1415          & 0.1434          & 0.1404          & 0.1458          \\
NSEI                 & \textbf{0.0485} & 0.0509 & 0.0862          & 0.0500          & 0.0496    & \textbf{0.0882} & 0.0912          & 0.0914          & 0.0896          & 0.0892    & 0.1413          & 0.1430          & 0.1685          & \textbf{0.1405} & 0.1407          \\
OMXC20               & \textbf{0.0422} & 0.044  & 0.0432          & 0.0426          & 0.0435    & 0.0834          & 0.0840          & 0.0842          & \textbf{0.0829} & 0.0838    & 0.1388          & 0.1376          & \textbf{0.1353} & 0.1367          & 0.1367          \\
OMXHPI               & \textbf{0.0454} & 0.0532 & 0.0529          & 0.0516          & 0.0460    & \textbf{0.0848} & 0.0944          & 0.0943          & 0.0926          & 0.0870    & \textbf{0.1373} & 0.1467          & 0.1451          & 0.1447          & 0.1417          \\
OMXSPI               & \textbf{0.0451} & 0.0551 & 0.0564          & 0.0532          & 0.0483    & \textbf{0.0885} & 0.0986          & 0.0981          & 0.0960          & 0.0921    & \textbf{0.1475} & 0.1537          & 0.1501          & 0.1504          & 0.1493          \\
OSEAX                & 0.0421          & 0.0437 & 0.0735          & \textbf{0.0418} & 0.0565    & 0.0845          & 0.0856          & 0.0851          & \textbf{0.0843} & 0.0989    & 0.1423          & \textbf{0.1406} & 0.1619          & 0.1413          & 0.1534          \\
RUT                  & 0.0559          & 0.0567 & 0.0574          & \textbf{0.0541} & 0.0717    & 0.1095          & 0.1117          & 0.1118          & \textbf{0.1085} & 0.1244    & 0.1818          & 0.1842          & 0.1866          & \textbf{0.1817} & 0.1936          \\
SMSI                 & 0.0497          & 0.0512 & \textbf{0.0496} & 0.0503          & 0.0541    & \textbf{0.0911} & 0.0936          & 0.0929          & 0.0913          & 0.0976    & \textbf{0.1466} & 0.1501          & 0.1488          & 0.1468          & 0.1521          \\
SPX                  & \textbf{0.0449} & 0.0487 & 0.0545          & 0.0477          & 0.0486    & \textbf{0.0867} & 0.0923          & 0.0924          & 0.0898          & 0.0925    & \textbf{0.1432} & 0.1472          & 0.151           & 0.1446          & 0.1482          \\
SSEC                 & \textbf{0.0474} & 0.05   & 0.0595          & 0.0490          & 0.0498    & 0.0887          & 0.0889          & 0.0903          & \textbf{0.0880} & 0.0890    & 0.1403          & 0.1392          & 0.1527          & \textbf{0.1389} & 0.1399          \\
SSMI                 & \textbf{0.0355} & 0.0368 & 0.0361          & 0.0355          & 0.0496    & \textbf{0.0705} & 0.0730          & 0.0747          & \textbf{0.0705} & 0.0837    & \textbf{0.1156} & 0.1204          & 0.1175          & 0.1179          & 0.1273          \\
STI                  & \textbf{0.0332} & 0.0355 & 0.035           & 0.0357          & 0.0369    & \textbf{0.0643} & 0.0650          & 0.0649          & 0.0651          & 0.0657    & 0.1045          & 0.1037          & \textbf{0.1033} & 0.1046          & 0.1043          \\
STOXX50E             & \textbf{0.0484} & 0.0547 & 0.059           & 0.0528          & 0.0660    & \textbf{0.0954} & 0.1019          & 0.1023          & 0.0987          & 0.1107    & \textbf{0.1537} & 0.1602          & 0.1613          & 0.1562          & 0.1692         \\    \hline 
\end{tabular}
 \end{table}
\centering{\tiny{Note: Bold numbers indicate the favoured models}}
\end{landscape}

The second criterion utilized to evaluate VaR forecasts is the VaR violation rate (VRate), which measures the proportion of returns during the forecast period that surpasses or equals the VaR forecast $\widehat{Q}_{t}$.
\begin{align}
    \textup{VRate }= \frac{1}{T_\text{test}}\sum_{t = T+1}^{T+T_\text{test}}I(r_{t}<\widehat{Q}_{t})
    \label{eq: 28}
\end{align}
where $T$ is the in-sample size and $T_\text{test}$ is the test sample size. Models with VRate close to $\alpha$, or equivalently $\frac{\textup{VRate}}{\alpha}$ close to 1, are preferred. Table \ref{table:4} shows the VRate values for the 31 markets for the three different $\alpha$ values we considered in this study.
The results displayed in Table \ref{table:4} highlight that the RNN-HAR model surpasses the other models in performance. Specifically, for $\alpha = 1\%$ and $\alpha = 2.5\%$, the proposed model is the best performer for more than $90\%$  of the markets. Likewise, for $\alpha = 5\%$, it maintains a lead in performance with a rate of $77\%$.

As the third performance measure, we use the "Diebold and Mariano test with Quandt-Andrews break" (DQ) out-of-sample test \cite{engle2004caviar}.
As pointed out by \cite{engle2004caviar}, the out-of-sample DQ test does not depend on the estimation procedure, but only the sequences of the VaR forecast and the corresponding portfolio
values are needed. 
In this test, a series
of ``Hits'' are calculated by $H_{t}=I(y_{t}<\text{VaR}_{t})-\alpha$
for the null hypothesis and iid series with rate $\alpha$. When the
null hypothesis is true, it can be shown that $\mathbb{E}(H_{t})=0$ and $\mathbb{E}(H_{t}W_{it})=0$
\citep{gerlach2016bayesianB} 
where $W$ has $q$ explanatory variables
that are in the information set at time $t-1$ when the forecast $\text{VaR}_{t}$
is made. To check whether all parameters in a regression of $H$ on
$W$ equal zero, the following DQ test statistic has been derived,
\begin{align}
DQ\left(q\right)=\frac{H^{'}W\left(W^{'}W\right)^{-1}W^{'}H}{\alpha\left(1-\alpha\right)}\sim \chi_{q}^{2}, \label{Eq:29}
\end{align}
Following \cite{engle2004caviar} 
and \cite{gerlach2016bayesianB}, 
we employ 4 lagged hits in this paper as follows.
\begin{align*}
W_{t}^{T}&=(1, H_{t-1}, \text{VaR}_{t})\;\;\; \text{denoted as } DQ_{1},\\
W_{t}^{T}&=(1, H_{t-1}, H_{t-2}, \text{VaR}_{t})\;\;\; \text{denoted as } DQ_{2},\\
W_{t}^{T}&=(1, H_{t-1}, H_{t-2}, H_{t-3}, \text{VaR}_{t})\;\;\; \text{denoted as } DQ_{3},\\
W_{t}^{T}&=(1, H_{t-1}, H_{t-2}, H_{t-3}, H_{t-4}, \text{VaR}_{t})\;\;\; \text{denoted as } DQ_{4}.
\end{align*}
Table \ref{table: 5} shows the DQ test results for each significance level we considered in this study. Minimum rejections are better. The table shows the number of markets for which the DQ tests reject each model. The proposed RNN-HAR model has the lowest number of rejections out of 31 markets considered for each significance level.

\setlength{\tabcolsep}{2pt}
\begin{landscape}
\mbox{}\vspace*{0.3cm} 
\begin{table}[H]
\mbox{}\hspace*{0.25cm} 
\caption{VRate } 
\centering{}{\tiny{}} \label{table:4}     
\scriptsize\begin{tabular}{|l|ccccc|ccccc|ccccc|}  
\multicolumn{1}{l}{} & \multicolumn{5}{c}{$\alpha = 1\%$}   & \multicolumn{5}{c}{$\alpha = 2.5\%$} & \multicolumn{5}{c}{$\alpha = 5\%$}                                                        \\   \hline

Market  & RNN-HAR         & HAR    & LevHAR          & SqrtHAR         & RHARGARCH & RNN-HAR         & HAR             & LevHAR          & SqrtHAR         & RHARGARCH & RNN-HAR         & HAR             & LevHAR          & SqrtHAR         & RHARGARCH       \\    \hline
AEX      & \textbf{1.40} & 2.90          & 3.30          & 2.50          & 3.10    & \textbf{1.40} & 1.72          & 1.80          & 1.68          & 1.88          & \textbf{1.26} & 1.32          & 1.66          & \textbf{1.26} & 1.38          \\
AORD     & \textbf{1.70} & 3.30          & 5.60          & 3.10          & 3.40    & \textbf{1.48} & 1.68          & 1.68          & 1.64          & 1.92          & \textbf{1.10} & 1.18          & 2.02          & 1.16 & 1.36          \\
BFX      & \textbf{1.00} & 2.30          & 2.80          & 2.10          & 2.70    & \textbf{1.08} & 1.68          & 1.64          & 1.56          & 2.04          & \textbf{1.18} & 1.24          & 1.22          & 1.22         & 1.36          \\
BSESN    & \textbf{1.80} & 2.50          & 6.40          & 2.30          & 2.50    & \textbf{1.36} & 1.76          & 1.80          & 1.60          & 1.68          & \textbf{1.16} & 1.20          & 2.10          & \textbf{1.16} & 1.22          \\
BVLG     & \textbf{0.76} & 1.50          & 1.88          & 2.90          & 1.89    & \textbf{0.81} & 1.77          & 1.40          & 1.72          & 1.46          & \textbf{0.91} & 1.15          & 1.13 & 1.19          & 1.21          \\
BVSP     & 1.90         & 2.10         & 2.50          & \textbf{1.80} & 2.10    & \textbf{1.32} & 1.52          & 1.44 & 1.52          & 1.36          & 1.12          & 1.16 & 1.22          & 1.12          & \textbf{1.08} \\
DJI      & \textbf{2.30} & 2.80          & 5.30          & 2.50          & 3.10    & \textbf{1.48} & 1.96          & 1.92         & 1.64          & 1.96          & \textbf{1.24} & 1.46          & 1.92          & 1.44          & 1.46          \\
FCHI     & \textbf{1.80} & 3.40          & 4.00          & 3.00          & 3.60    & \textbf{1.32} & 2.00          & 1.96          & 1.88          & 2.08          & \textbf{1.20} & 1.44          & 1.54          & 1.40          & 1.44          \\
FTMIB    & \textbf{1.50} & 2.90          & 3.30          & 2.90          & 3.00    & \textbf{1.16} & 1.64          & 1.72          & 1.60 & 1.72          & \textbf{0.96} & 1.18          & 1.22          & 1.16          & 1.16          \\
FTSE     & \textbf{2.50} & 3.30          & 3.60          & 3.00         & 3.90    & \textbf{1.48} & 1.76          & 1.76          & 1.76          & 2.24          & \textbf{1.18} & 1.32          & 1.60          & 1.20          & 1.48          \\
GDAXI    & \textbf{2.20} & 4.00          & 4.60          & 3.60          & 3.20    & \textbf{1.64} & 2.52          & 2.56          & 2.44          & 2.08          & \textbf{1.12} & 1.86          & 1.74          & 1.74          & 1.64          \\
GSPTSE   & \textbf{1.40} & 3.30         & 6.80          & 3.10          & 3.40    & \textbf{1.24} & 2.00         & 1.96          & 2.00          & 1.96          & \textbf{1.10} & 1.38          & 2.50          & 1.34          & 1.40          \\
HIS      & \textbf{1.30} & 2.60          & 2.40          & 2.60          & 2.80    & \textbf{1.28} & 1.68          & 1.64         & 1.72          & 1.76          & \textbf{1.26} & 1.32          & 1.56          & \textbf{1.26} & 1.28          \\
IBEX     & \textbf{1.50} & 1.90          & 2.00 & 2.00          & 2.30    & 1.24          & \textbf{1.16} & 1.24          & \textbf{1.16} & 1.48 & 0.88          & 0.86          & \textbf{0.98} & 0.88 & 0.96          \\
IXIC     & \textbf{1.40} & 2.10          & 2.40          & 2.00          & 2.40    & \textbf{1.24} & 1.36          & 1.40          & 1.36 & 1.68          & \textbf{1.12} & 1.32          & 1.44          & 1.28 & 1.42          \\
KS11     & \textbf{0.90} & 2.20          & 2.10 & 1.90          & 2.30   & \textbf{0.80} & 1.84          & 1.84          & 1.60          & 1.68          & \textbf{1.14} & 1.54         & 1.52 & 1.38          & 1.44          \\
KSE      & \textbf{1.30} & 2.90         & 2.60          & 2.90         & 2.70    & \textbf{1.20} & 1.96          & 2.12          & 1.92          & 1.92          & \textbf{1.40} & 1.52          & 1.50          & 1.50         & \textbf{1.40} \\
MXX      & 1.50          & \textbf{1.10} & 1.50 & 1.20          & 2.80   & 1.24          & \textbf{1.00} & 1.04         & 0.88 & 1.84         & 1.36          & 0.86          & \textbf{0.90} & 0.84          & 1.38          \\
N225     & \textbf{1.00} & 2.20         & 2.80          & 2.00         & 2.00    & \textbf{1.00} & 1.60         & 1.72         & 1.48          & 1.32          & \textbf{1.06} & 1.34          & 1.34          & 1.26        & 1.12          \\
NSEI     & \textbf{2.00} & 2.60          & 8.00         & 2.50          & 2.50   & \textbf{1.32} & 1.48          & 1.56         & 1.48          & 1.52          & \textbf{1.12} & 1.14          & 2.34          & \textbf{1.12} & 1.24         \\
OMXC20   & \textbf{1.30} & 1.90          & 2.20          & 1.90          & 2.00    & \textbf{1.12} & 1.28          & 1.32          & 1.28 & 1.32          & \textbf{1.02} & 1.08          & 1.32 & 1.06          & 1.10          \\
OMXHPI   & \textbf{1.50} & 3.70          & 4.00         & 3.60          & 1.80    & \textbf{1.32} & 2.24         & 2.32          & 2.16          & 1.56          & \textbf{1.14} & 1.64         & 1.72         & 1.58          & 1.20         \\
OMXSPI   & \textbf{2.00} & 3.80          & 4.40          & 3.50          & 2.50   & \textbf{1.32} & 2.28          & 2.32          & 2.28          & 1.92         & \textbf{1.16} & 1.60          & 1.70         & 1.58          & 1.36          \\
OSEAX    & \textbf{1.40} & 2.20         & 5.20         & 2.00 & 2.60    & \textbf{1.12} & 1.24          & 1.16          &1.36 & 1.48          & 1.12          & \textbf{1.04} & 1.82          & 1.08          & 1.08         \\
RUT      & 1.30          & 1.60         & 2.10         & \textbf{1.20} & 2.50    & 1.24         & 1.16          & 1.20          & \textbf{1.00} & 1.76 & \textbf{1.02} & 0.94          & 1.10          & 0.86 & 1.46          \\
SMSI     & \textbf{1.30} & 1.70         &2.20 & 1.70        & 2.70   & \textbf{1.00} & 1.32         & 1.32         & 1.08         & 1.52          & \textbf{0.98} & 0.92          & 1.18         & 0.96          & \textbf{1.02} \\
SPX      & \textbf{2.10} & 3.20          & 4.80         & 2.70          & 3.50    & \textbf{1.36} & 1.96          & 2.08          & 1.80         & 2.00          & 1.36& 1.40         & 2.06          & \textbf{1.32} & 1.48          \\
SSEC     & \textbf{0.80} & 2.30          & 4.30         & 2.10          & 2.40    & \textbf{1.00} & 1.40          & 1.40          & 1.32 & 1.48          & 1.12          & 1.16          & 1.88          & \textbf{1.04} & 1.16        \\
SSMI     & \textbf{1.30} & 2.60          & 2.80         & 2.00          & 3.30    & \textbf{1.04} & 1.60          & 1.68         & 1.44 & 1.56         & \textbf{1.14} & 1.24         & 1.26          & 1.20          & 1.20          \\
STI      & \textbf{1.20} & 1.60          & 1.70         & 1.50          & 2.20    & \textbf{1.12} & 1.12          & \textbf{1.12} & 1.16          & 1.36          & 1.28          & 0.90         & 0.90 & \textbf{0.94} & 1.12         \\
STOXX50E & \textbf{1.80} & 3.20          & 3.90          & 2.80          & 3.20    & \textbf{1.44} & 1.92          & 1.92         & 1.80          & 1.92          & \textbf{1.16} & 1.32         & 1.42          & 1.26         & 1.34         \\    \hline 
\end{tabular}
 \end{table}
\centering{\tiny{Bold numbers indicate the favoured models}}
\end{landscape}

\setlength{\tabcolsep}{2pt}
\begin{table}[H]
\mbox{}\hspace*{0.25cm} 
\caption{DQ test results}
\centering{}{\tiny{}}  \label{table: 5} 
\begin{tabular}{|l|cccc|cccc|cccc|}
\multicolumn{1}{c}{} \\ \hline & \multicolumn{4}{c} 
{$\alpha = 1\%$}  \vline & 
\multicolumn{4}{c}{$\alpha = 2.5\%$} \vline & \multicolumn{4}{c}{$\alpha = 5\%$} \vline  \\ \hline
 & DQ1    & DQ2    & DQ3    & DQ4    & DQ1    & DQ2    & DQ3   & DQ4   & DQ1    & DQ2    & DQ3   & DQ4   \\ \hline
RNN HAR              & \textbf{8}    & \textbf{8}    & \textbf{14}    & \textbf{14}    & \textbf{7}      & \textbf{5}     & \textbf{8}      & \textbf{7}      & \textbf{4}      & \textbf{3}      & \textbf{4}     & \textbf{4}     \\
HAR                 & 25     & 25     & 26    & 25     & 23     & 27     & 25     & 24     & 16     & 15     & 16    & 16    \\
LevHAR              & 28     & 28     & 29    & 29 & 21     & 26     & 22     & 21         & 19     & 20     & 20    & 20    \\
SqrtHAR               & 21     & 20     & 22    & 23  & 21     & 21     & 20     & 22        & 14     & 15   & 15    & 15  \\ 
RHARGARCH            & 28     & 27     & 28     & 28     & 25     & 23     & 25    & 26    & 21     & 20   & 20    & 19   \\\hline 
\end{tabular} 
\caption*{Bold numbers indicate the favoured models}
 \end{table}

Finally, we use the tail loss ratio to evaluate the effectiveness of the VaR forecasting methods. VaR estimates the maximum potential loss a portfolio or investment may incur over a specified time horizon at a given confidence level. However, VaR alone may not adequately capture extreme losses or tail risk scenarios of significant concern to investors and risk managers. The Tail Loss Ratio addresses this limitation by focusing on the losses beyond the VaR threshold, often called ``tail losses". It quantifies the severity of these extreme losses relative to the VaR estimate. We use the following equation to calculate the tail loss ratio
\begin{align}  \label{eq: 30}
        \textrm{Tail loss ratio} = \frac{\sum_{t=1}^{T_\text{test}}\max(0,y_{t}-\textup{VaR}_{t})}{\sum_{t=1}^{T_\text{test}}y_{t}},
\end{align}
where $y_{t}$ are the returns in the forecast period. The model with the lowest tail loss ratio is generally considered the best performer. Table \ref{table:6} lists the tail loss ratios for the different markets and significant levels. For $\alpha = 1\%$, our proposed model is better than the other models in 61\% of the markets, and for $\alpha = 2.5\%$, $\alpha = 5\%$, it is $55\%$,  $48\%$ respectively.

\setlength{\tabcolsep}{2pt}
\begin{landscape}
\mbox{}\vspace*{0.8cm} 
\begin{table}[H]
\mbox{}\hspace*{0.35cm} 
\caption{Tail loss ratio } 
\centering{}{\tiny{}} \label{table:6}     
\tiny\begin{tabular}{|l|ccccc|ccccc|ccccc|}  
\multicolumn{1}{l}{} & \multicolumn{5}{c}{$\alpha = 1\%$}   & \multicolumn{5}{c}{$\alpha = 2.5\%$} & \multicolumn{5}{c}{$\alpha = 5\%$}                  \\   \hline

Market  & RNN-HAR         & HAR    & LevHAR          & SqrtHAR         & RHARGARCH & RNN-HAR         & HAR             & LevHAR          & SqrtHAR         & RHARGARCH & RNN-HAR         & HAR             & LevHAR          & SqrtHAR         & RHARGARCH       \\    \hline
AEX                        & 193.42    & 169.60          & \textbf{160.92}  & 170.55           & 165.17          & 154.70    & 144.21           & 144.26           & 144.95            & \textbf{140.51}  & 122.63  & 122.76           & \textbf{116.74}  &123.33   & 119.72           \\
AORD                       & \textbf{-493.65}   & -417.16         & -384.59          & -416.23          & -383.90         & \textbf{-383.58}   & -354.10          & -353.56          & -353.26          & -326.53          & \textbf{-319.27} & -300.49          & -279.00         &-299.74  & -278.08          \\
BFX                        & \textbf{-453.58}   & -394.35         & -384.42          & -393.40          & -371.20         & \textbf{-359.75}   & -334.60          & -334.95          & -333.66           & -315.38          & \textbf{-288.36} & -284.17          & -277.17          & -283.30           & -268.51          \\
BSESN                      & \textbf{-2448.50}  & -2115.00        & -1942.60         & -2114.50         & -2146.10         & \textbf{-1970.30}  & -1797.40        & -1815.10         & -1796.10         & -1823.00        & \textbf{-1630.40}   & -1529.70         & -1413.50         & -1528.10 & -1550.50         \\
BVLG                       & 179.73    & 151.40          & 149.20           & 149.65           & \textbf{145.19} & 150.28    & 126.16           & 126.57           & 127.19            & \textbf{123.30}  & 119.39  & 109.13           & 107.53  & 108.21            & \textbf{104.91}  \\
BVSP                       & \textbf{-279.13}   & -270.55         & -266.81          & -272.10 & -272.17         & \textbf{-239.76}   & -229.33         & -229.02 & -230.76           & -230.77         & \textbf{-200.38} & -194.64 & -192.32          & -195.88           & -195.92 \\
DJI                        & 533.38    & 472.58          & \textbf{450.85}  & 478.48           & 459.93          & 441.18    & 402.26           & 401.94           & 406.95            & \textbf{391.80}  & 359.35  & 343.67           & \textbf{329.00}  & 347.00            & 334.68           \\
FCHI                       & 331.70    & 282.12          & \textbf{271.95}  & 283.62           & 285.49          & 268.39    & 240.11          & \textbf{239.65}  & 241.23            & 242.88           & 220.85  & 204.76          & \textbf{197.73}  & 205.57            & 206.96           \\
FTMIB                      & \textbf{-1720.20}  & -1522.00        & -1487.10         & -1531.80         & -1520.70        & \textbf{-1437.60}  & -1293.20         & -1288.30        & -1301.30 & -1292.20        & \textbf{-1162.30}   & -1099.50         & -1075.60         & -1106.00          & -1098.80         \\
FTSE                       & \textbf{-400.66}   & -362.46         & -344.20          & -364.60          & -321.08        & \textbf{-333.16}   & -308.00          & -307.24          & -309.75           & -273.83          & \textbf{-273.49} & -261.80          & -249.41          & -263.21           & -233.91          \\
GDAXI                      & \textbf{-524.68}   & -416.16         & -408.63         & -420.14          & -443.58         & \textbf{-415.46}   & -354.83          & -355.61          & -358.01           & -377.41          & \textbf{-351.61} & -303.40          & -298.20          & -305.94           & -322.07          \\
GSPTSE                     & \textbf{-1210.40}  & -919.52         & -838.98          & -916.97          & -904.56         & \textbf{-941.13}   & -783.44          & -785.06          & -781.67           & -772.26          & \textbf{-757.73} & -669.58          & -616.71          & -668.34           & -661.19          \\
HIS                        & \textbf{-88.87}    & -72.89          & -72.14           & -73.94           & -72.86          & \textbf{-66.30}    & -61.69           & -61.57          & -62.56            & -61.68           & \textbf{-54.87}  & -52.30           & -51.84           & -53.01   & -52.28           \\
IBEX                       & \textbf{-318.57}   & -297.63         & -292.87 & -296.08          & -278.15         & \textbf{-253.57}   & -252.10 & -252.62          & -250.88  & -236.20 & \textbf{-214.61} & -213.62          & -210.42 & -212.57  & -200.51          \\
IXIC                       & 168.95    & 139.74          & 137.15           & 140.40           & \textbf{132.74} & 127.54    & 118.69           & 118.78           & 119.23   & \textbf{112.87}  & 109.46    & 101.03           & 99.33            &101.41   & \textbf{96.23}   \\
KS11                       & \textbf{-205.20}   & -161.25         & -159.84 & -164.73          & -159.91         & \textbf{-161.70}   & -136.68          & -136.41          & -139.51           & -135.58          & \textbf{-127.83}   & -116.04          & -114.98 & -118.30           & -115.09          \\
KSE                        & \textbf{-49.02}    & -38.37          & -37.78           & -38.27           & -39.42          & \textbf{-36.46}    & -32.45           & -32.46           & -32.37            & -33.32           & \textbf{-28.48}  & -27.49           & -27.08           & -27.42            & -28.20  \\
MXX                        & \textbf{-90.22}    &-86.76 & -86.35  & -88.90           & -73.34          & -69.91             & -73.29  & -72.85           & \textbf{-75.07}   & -62.15           & -52.41           & -61.87           & -61.63  & \textbf{-63.34}   & -52.74           \\
N225                       & 254.26    & 204.36          & \textbf{200.27}  & 205.75           & 220.35          & 195.81    & \textbf{173.42}  & 173.20           & 174.55            & 186.79           & 157.28  & 147.44           & \textbf{144.67}  & 148.33            & 158.44           \\
NSEI                       & \textbf{-819.67}   & -764.76         & -684.75          & -767.44          & -751.25         & \textbf{-698.66}   & -649.48          & -653.64          & -651.41           & -637.98          & \textbf{-573.04} & -551.97          & -498.18         & -553.20  & -542.38          \\
OMXC20                     & 170.96    & 150.75          & \textbf{146.29}  & 153.49           & 151.16          & 140.74    & 127.92           & \textbf{127.75}  & 130.20   & 128.28           & 116.48  & 108.66           & \textbf{105.54}  & 110.52            & 108.91           \\
OMXHPI                     & \textbf{-23893.00} & -19177.00       & -18721.00        & -19084.00        & -22304.00         & \textbf{-20062.00} & -16336.00        & -16389.00        & -16255.00         & -18919.00        & \textbf{-16512.00}    & -13945.00        & -13623.00        & -13874.00         & -16066.00        \\
OMXSPI                     & 536.91    & 446.58          & \textbf{428.12}  & 444.37           & 499.93          & 454.30    & 380.51           & 381.29           & \textbf{378.49}   & 424.61           & 376.17  & 324.87           & \textbf{311.52}  & 322.99            & 361.11           \\
OSEAX                      & \textbf{-334.33}   & -308.47         & -270.29          & -307.11 & -304.46          & \textbf{-275.07}   & -261.22          & -262.44          & -260.12  & -258.19          & \textbf{-222.99} & -221.21 & -195.42          & -220.50           & -218.95          \\
RUT                        & -216.71            & -214.59         & -211.72          & \textbf{-216.80} & -180.65         & -178.43           & -181.60          & -181.73          & \textbf{-183.37}  & -153.46 & -149.31 & -153.65          & -151.82          & \textbf{-155.12}  & -130.59          \\
SMSI                       & \textbf{-261.68}   & -242.74         & -235.85 & -239.28          & -223.51         & \textbf{-215.18}   & -205.58          & -206.46          & -202.58           & -189.77          &-170.63 & \textbf{-174.19} & -169.51          & -171.62           & -161.16 \\
SPX                        & 185.95    & 155.51         & \textbf{149.27}  & 157.34           & 152.20          & 148.06    & 132.57           & 132.37           & 133.98            & \textbf{129.84}  & 119.48   & 113.25           & \textbf{109.01}  & 114.36   & 111.06           \\
SSEC                       & \textbf{-381.58}   & -287.10         & -272.88          & -296.19          & -287.48         & \textbf{-273.34}   & -243.51          & -242.94          & -251.05  & -243.86          & -211.47          & -206.60          & -197.76          & \textbf{-212.82}  & -206.95          \\
SSMI                       &150.93    & 127.59          & 122.58           & 129.65           & \textbf{119.93}  & 121.08    & 108.48           & 108.26           & 110.11   & \textbf{102.13}  & 94.13   & 92.38            & 88.84            & 93.68             & \textbf{87.08}   \\
STI                        & \textbf{-211.59}   & -189.02         & -189.71          & -188.72          & -178.44          & -160.09   & -159.99          & \textbf{-160.19} & -159.74           & -151.16          & -124.08          & -135.38          & \textbf{-135.83} & -135.25  & -128.16          \\
STOXX50E                   &399.62    & 342.30          & \textbf{330.25}  & 343.17           & 335.47          & 317.63    & 291.38           & 292.62           & 291.88            & \textbf{285.62}  & 259.24  & 248.20           & \textbf{239.79}  & 248.44            & 243.74   \\    \hline 
\end{tabular}
 \end{table}
\centering{\tiny{Bold numbers indicate the favoured models}}
\end{landscape}


\section{Conclusion} 
Our research expands the HAR model by integrating RNN structures that calculate daily, weekly and monthly non-linear and long-term effects of realized variances on VaR. The new RNN-HAR model directly estimates VaR and uses quantile scores to avoid assumptions about the return distribution, thereby incorporating return series and avoiding potential inaccuracies from relying solely on realized measures.

We use SMC with likelihood annealing for in-sample analysis and SMC with data annealing for out-of-sample forecasting. Our extensive empirical study covers 31 major stock markets, demonstrating that the proposed RNN-HAR model outperforms other HAR extensions across all significance levels we considered.

Looking forward, a promising avenue for future research involves further enriching the model with Long Short-Term Memory structures and incorporating multiple realized measures. These advancements hold potential for achieving even higher levels of predictive accuracy and robustness in financial forecasting.



\bibliographystyle{apalike}
\bibliography{referencelist}
\end{document}